# Analyzing Large Biological Datasets with an Improved Algorithm for MIC

Shuliang Wang, Yiping Zhao

School of Software, Beijing Institute of Technology, Beijing 100081, China

**Abstract:** A computational framework utilizes the traditional similarity measures for mining the significant relationships in biological annotations is recently proposed by Tatiana V. Karpinets et al. [2]. In this paper, an improved approximation algorithm for MIC (maximal information coefficient) named IAMIC is suggested to perfect this framework for discovering the hidden regularities between biological annotations. Further, IAMIC is the enhanced algorithm for approximating a novel similarity coefficient MIC with generality and equitability, which makes it more appropriate for data exploration. Here it is shown that IAMIC is also applicable for identify the associations between biological annotations.

***Keywords:*** *IAMIC*, *biological annotations, data exploration*

1. INTRODUCTION

With the amount of biological data dramatic increases, more and more research community will be certainly in the process of working on annotations [1]. Many biological databases is currently collecting the annotation information of different biological objects at amazing rate, which requires appropriate analysis tools to put in place. Tatiana V. Karpinets et al. [2] recently proposed a novel approach for mining modular structure, relationships and regularities in lager biological datasets. This method applied the similarity measures, such as Pearson correlation coefficient to discover the hidden associations between pairwise annotations.

Here we introduce a new measure to identify the relationship between pairs of annotations. The new measure is called IAMIC, the improved approximation algorithm for maximal information coefficient, which based on the algorithm of Reshef et al. [3], and it is an enhanced solution for detecting the relationships between variables in large data sets with better tradeoffs between equitability and time. In our work, we present a comparison between the proposed algorithm and the traditional similarity measures to uncover novel regularities between biological annotations.

The rest of this paper is organized as follows. In section 2, the work that has been done previously in this field is summarized. In section 3, the description of applying IAMIC in the biological annotations is presented. In section 4, the experiment results are compared to the traditional similarity algorithms used by Tatiana V. Karpinets et al. [2].

## 2. RELATED WORK

To satisfy the increasing requirement of computational methods that process collected biological information into new knowledge, Tatiana V. Karpinets et al. [2] suggested an approach for analyzing the records of collected biological annotations in order to discover the associated annotations. This approach firstly converted the original annotations into the type-value formatted transactions which were conductive for preserving the semantic structure of data, and then calculated the records of annotations co-occurrence in the transactions to formulate a support matrix in which each row or column shown the co-occurrence values between a specific annotation and other annotations. Further, it used the one of the traditional similarity measures Pearson correlation coefficient to mine the relationship of annotations.

Tatiana V. Karpinets et al. [2] applied the method to the GOLD datasets and demonstrated the ability to identify the significant relationships between biological annotations that especially do not co-occur with each other. Particularly, they discussed it may had some limitations as other statistical analysis shown, and then expected novel measures of similarity, such as maximal information coefficient (MIC) [3] could be utilized to discover hidden regularities in collected datasets.

MIC is an interesting statistical measure proposed by Reshef et al. [3], which has always been attracting the eye of many fields due to its two properties: generality and equitability [4][5]. Compared to traditional similarity measures like Pearson correlation coefficient, MIC was more appropriate for a wide range of associations, not limited to specific relationships, and more equitable for no preference to certain functional types. MIC has been applied to biological terms successfully [6] [7] [8], including clinical data, genomics and virology applications.

Although MIC was a great advance, it remained some drawbacks as presented in [9] [10] [11] and was argued less practical than distance correlation [13] [14] and HHG [15]. And then the later study of Reshef et al. [12] suggested that only the approximation algorithm for MIC resulted in these existed limitations rather than the intrinsic feature of MIC. Furthermore, we have been working on the improvement of the standard approximation algorithm for MIC to come more close to the true value of MIC. In this work, we want to apply the improved algorithm for MIC called IAMIC to the framework proposed by Tatiana V. Karpinets et al. [2] to uncover novel associations between biological annotations.

## 3. MINEING BIOLOGICAL DATASETS WITH IAMIC

Here we are going to describe the main steps of mining biological annotations with an improved algorithm for calculating MIC named IAMIC. Firstly, we will introduce the principle of this algorithm for better understanding of whole work in our paper.

**3.1 THE PRINCIPLE OF IAMIC**

Reshef et al.[3] stated that MIC is the metric value for the relationships between two variables. And he provided a MIC calculating algotithm, like a fitting method, which encapsulates the 2D splashes with grid partition. In other words, MIC is the value related to the grid partition that best reflects the true relationship between two variables. This fitting like grid partition method has the property of equitability naturally. It means that MIC will give similar scores to different functional relationships with similar noise levels or similar $R^2$ (coefficient of determination).

To calculate the MIC value, the standard algorithm proposed by Reshef et al. [3] simply equipartition on y-axis and then search the optimal grid partition on x-axis. That algorithm seems to reveal the most possible relationship between two variables. But in fact it is just an approximation for real MIC value. And that approximation will lead to a wrong answer. The simplest way to get the real MIC value is to enumerate all the possible grid partition, and find the best fitting one. However, the time expend is too high to follow. Wang et al. [16] proposed a fast and high accuracy method to approximate the real MIC value called IAMIC. This new algorithm attempts to find a better partition on y-axis through quadratic optimization instead of violence search.

Specifically, IAMIC utilizes quadratic optimization on the largest value of each row of the characteristic matrix calculated by original algorithm in [3]. It retains the generality and improves the equitability in MIC, thus it is also more appropriate for mining the novel associations hidden in collected datasets than other similarity measures like Pearson correlation coefficient which has limitations for nonlinear relationships.

## 3.2 THE SIMPLE WORKFLOW OF APPLICATION

After brief introduction about new algorithm IAMIC, it can more clearly present the workflow of mining biological annotations with this method.

In our work, we will adopt the preprocessing process of datasets introduced by Tatiana V. Karpinets et al. [2]. It transfers the table formatted records to type-value formatted annotations. For each unique annotation, we calculate the values of co-occurrence with all other annotations in input datasets in order to build one row or column of a support matrix which is the object for later analysis by IAMIC. And then make use of new similarity measure IAMIC to quantify the associations between every pairwise annotation by computing their co-occurrence values recorded in the support matrix. Finally use the biological information visualization tool Cytoscape to show the data analysis results. The details of this workflow are presented as Figure 1.

Work flow: (a) –step1→ (b) –step2→ (c) –step3→ (d)

| ID | Super-kingdom | Group | Gram-stain | Shape |
|---|---|---|---|---|
| 1 | Bacteria | Firmicutes | + | Cocci |
| 2 | Bacteria | Firmicutes | - | Rod |
| 3 | Bacteria | Alphaproteobacteria | - | Rod |
| 4 | Bacteria | Betaproteobacteria | - | Rod |
| 5 | Bacteria | Other_Bacteria | - | Spiral |

(a) Collected table records

**ID1:** { Super-kingdom: Bacteria , Group: Firmicutes , Gram-stain:+ , Shape: Cocci }
**ID2:** { Super-kingdom: Bacteria , Group: Firmicutes , Gram-stain:- , Shape: Rod }
**ID3:** { Super-kingdom: Bacteria , Group: Alphaproteobacteria , Gram-stain:- , Shape: Rod }
**ID4:** { Super-kingdom: Bacteria , Group: Betaproteobacteria , Gram-stain:- , Shape: Rod }

**ID5:**{ Super-kingdom: Bacteria , Group: Other- bacteria , Gram-stain:- , Shape: Spiral }

(b) Type-value formatted records

|   | 1 | 2 | 3 | 4 | 5 | 6 | 7 | 8 | 9 | 10 |
|---|---|---|---|---|---|---|---|---|---|----|
| 1 | 5 | 2 | 1 | 1 | 1 | 1 | 4 | 1 | 3 | 1 |
| 2 | 2 | 2 | 0 | 0 | 0 | 1 | 1 | 1 | 1 | 0 |
| 3 | 1 | 0 | 1 | 0 | 0 | 0 | 1 | 0 | 1 | 0 |
| 4 | 1 | 0 | 0 | 1 | 0 | 0 | 1 | 0 | 1 | 0 |
| 5 | 1 | 0 | 0 | 0 | 1 | 0 | 1 | 0 | 0 | 1 |
| 6 | 1 | 1 | 0 | 0 | 0 | 1 | 0 | 1 | 0 | 0 |
| 7 | 4 | 1 | 1 | 1 | 1 | 0 | 4 | 0 | 3 | 1 |
| 8 | 1 | 1 | 0 | 0 | 0 | 1 | 0 | 1 | 0 | 0 |
| 9 | 3 | 1 | 1 | 1 | 0 | 0 | 3 | 0 | 3 | 0 |
| 10 | 1 | 0 | 0 | 0 | 1 | 0 | 1 | 0 | 0 | 1 |

| Super-kingdom: Bacteria | 1 |
| Group: Firmicutes | 2 |
| Group: Alphaproteobacteria | 3 |
| Group: Betaproteobacteria | 4 |
| Group: Other_Bacteria | 5 |
| Gram-stain:+ | 6 |
| Gram-stain:- | 7 |
| Shape: Cocci | 8 |
| Shape: Rod | 9 |
| Shape: Spiral | 10 |

(c) Support matrix

|   | 1 | 2 | 3 | 4 | 5 | 6 | 7 | 8 | 9 | 10 |
|---|---|---|---|---|---|---|---|---|---|----|
| 1 | — | 0.42 | 0.56 | 0.56 | 0.32 | 0.14 | 0.88 | 0.14 | 0.88 | 0.32 |
| 2 | 0.42 | — | 0.05 | 0.05 | 0.02 | 0.42 | 0.28 | 0.42 | 0.28 | 0.02 |
| 3 | 0.56 | 0.05 | — | 0.26 | 0.02 | 0.05 | 0.56 | 0.05 | 0.56 | 0.02 |
| 4 | 0.56 | 0.05 | 0.26 | — | 0.02 | 0.05 | 0.56 | 0.05 | 0.56 | 0.02 |
| 5 | 0.32 | 0.02 | 0.02 | 0.02 | — | 0.05 | 0.32 | 0.05 | 0.09 | 0.97 |
| 6 | 0.14 | 0.42 | 0.05 | 0.05 | 0.05 | — | 0.32 | 0.97 | 0.02 | 0.05 |
| 7 | 0.88 | 0.28 | 0.56 | 0.56 | 0.32 | 0.32 | — | 0.32 | 0.88 | 0.32 |
| 8 | 0.14 | 0.42 | 0.05 | 0.05 | 0.05 | 0.97 | 0.32 | — | 0.02 | 0.05 |
| 9 | 0.88 | 0.28 | 0.56 | 0.56 | 0.09 | 0.02 | 0.88 | 0.02 | — | 0.09 |
| 10 | 0.32 | 0.02 | 0.02 | 0.02 | 0.97 | 0.05 | 0.32 | 0.05 | 0.09 | — |

| Super-kingdom: Bacteria | 1 |
| Group: Firmicutes | 2 |
| Group: Alphaproteobacteria | 3 |
| Group: Betaproteobacteria | 4 |
| Group: Other_Bacteria | 5 |
| Gram-stain:+ | 6 |
| Gram-stain:- | 7 |
| Shape: Cocci | 8 |
| Shape: Rod | 9 |
| Shape: Spiral | 10 |

(d) Association matrix

Figure 1: workflow of applying IAMIC for mining biological annotations

In Figure 1, we present the workflow through simple example with four annotation types and five records corresponding to them. Step 1 shows the process of converting the collected table records (Figure 1(a)) to specific type-value formatted records (Figure 1(b)), and then generates a support matrix (Figure 1(c)) for unique annotations as shown in step 2. For instance, the value in the first row and second column of the support matrix refers to the number of records where annotation super_kingdom : Bacteria and annotation Group : Firmicutes co-occur, and especially, the values in the diagonal denote the number of records where annotations co-occur with itself, which are equal to the number of times in collected table recording such annotations. Step 3 uses IAMIC measure to estimate the similarity pairs of annotations by the support matrix produced in step 2 to create an association matrix (Figure 1(d)).

## 4. RESULTS AND COMPARISONS

We apply the method demonstrated in Chapter 3 to analyze the biological annotations document that is sample datasets in program provided by Tatiana V. Karpinets et al. [2]. Simultaneously, we also use the traditional similarity measures such as Pearson correlation coefficient, Spearman's rank correlation coefficient or Jaccard coefficient respectively to mining the same datasets.

In this section, we are going to reveal these comparision results as shown in Figure 2. The results are generated by visualization tool Cytoscape with the association matrix, of which the vertexes refer to the annotations and the edges refer to associations estimated by similarity values between the annotations.

There are 6782 collected table records, which contain 1109 unique annotations. We select 308 high frequency annotations that are found more than 6 records in the database during the experiments and screen out 2304 significant associations by setting p-value threshold 0.05. The comparision results from Figure 2 apparently show some advantages for IAMIC. Firstly, the measure IAMIC has better cluster feature than others. And some tight relationships between annotations are closer, unremarkable relations are far away, which bring us a direct information about which annotations are more likely to coexist. That is useful for analyzing the biological annotations. Secondly, it is easy to find that IAMIC lead to less isolated notation. That means IAMIC could mine the deeper relationship, which other method may not. Third, the more similar length between two annotations means a more disinterested measurement. That is to say, no matter what kind of relationships are, linear, curve, or even un-functional, IAMIC reveals their real relevance degree.

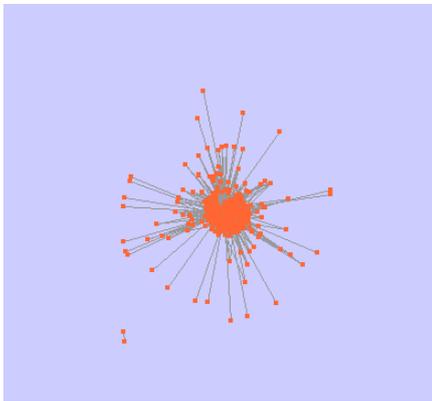　　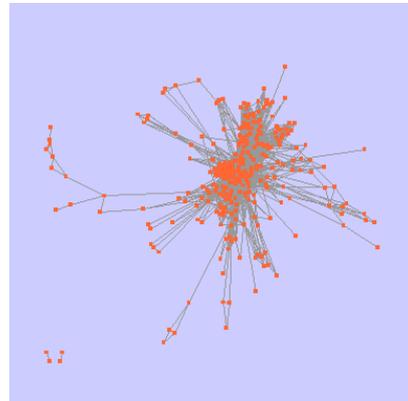

(a)  IAMIC　　　　　　　　　　　　(b)  Pearson correlation coefficient

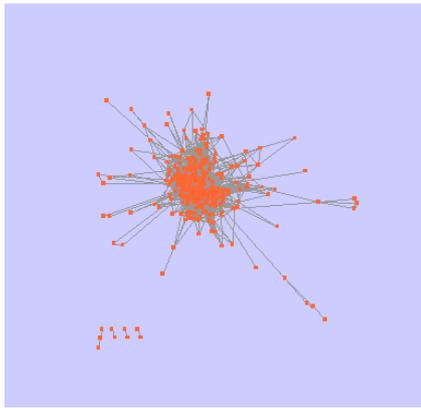 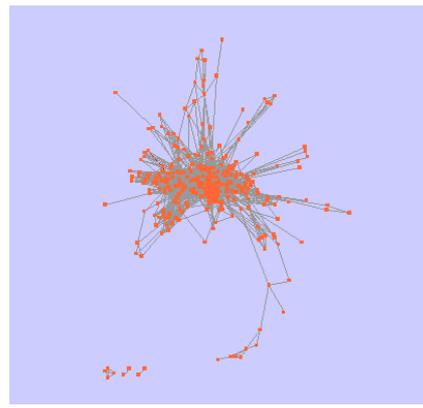

(c) Spearman's rank correlation coefficient      (d) Jaccard(cosine) coefficient

Figure 2: The comparision of analysis results

## 5. CONCLUSIONS

In this paper, we prefer to use our improved approximation algorithm for MIC called IAMIC to mine the biological annotations with the work provided by Tatiana V. Karpinets et al. [2]. We believe IAMIC is the novel similarity measure which takes advantages of data exploration than traditional approaches of similarity, and verified it is also applicative to biological annotations experimentally.

Finally, we hope a more appropriate similarity measure will be proposed in our next research, and it can be more conductive to uncover the hidden associations between increasing biological information.

**Acknowledgements**

This work was supported by National Natural Science Fund of China (61173061, 71201120), and Specialized Research Fund for the Doctoral Program of Higher Education (20121101110036).